\documentclass[%
 aip,
 jmp,%
 amsmath,amssymb,
 reprint,%
]{revtex4-1}

\usepackage{graphicx}
\usepackage{dcolumn}
\usepackage{bm}

\begin{document}

\preprint{AIP/123-QED}

\title{Appearance of Negative Differential Conductivity in Graphene Nanoribbons at High-Harmonics}
\thanks{Corresponding author. Email: rabpeace10gh@gmail.com.}

\author{M. Rabiu}
 \affiliation{Department of Applied Physics, Faculty of Applied Sciences, University for Development Studies, Navrongo Campus, Ghana.}
\author{S. Y. Mensah}%
 \email{profsymensah@yahoo.co.uk}
\affiliation{Department of Physics, Laser and Fiber Optics Center, University of Cape Coast, Cape Coast, Ghana.
}%

\author{S. S. Abukari}
\affiliation{Department of Physics, Laser and Fiber Optics Center, University of Cape Coast, Cape Coast, Ghana.
}%

\date{\today}

\begin{abstract}
We theoretically study current dynamics of graphene nanoribbons subject to bias dc and ac driven fields. We showed that graphene nanoribbons exhibit negative high-harmonic differential conductivity. Negative differential conductivity appears when bias electric filed is in the neighborhood of applied ac filed amplitude. We also observe both even and odd high-harmonic negative differential conductivity at wave mixing of two commensurate frequencies. The even harmonics are more pronounced than the odd harmonics. A possible use of the present method for generating terahertz frequencies at even harmonics in graphene is suggested.
%
\end{abstract}

\maketitle

\section{Introduction}

Graphene has continued to surprise scientists since its discovery in 2004 by Geim and his team \cite{KSnovoselov}. Theoretically, the carrier transport properties are fantastic. Especially, its high carrier mobility of 44000$cm^2V^{-1}s^{-1}$ \cite{RSShishir}. But attempts to utilize these astonishing properties in graphene devices are posing some difficulties. The limitation is probably due to several factors including; lack of band gap in graphene sheets, edge defects, disorder, among others. To overcome some of these obstacles, particularly the lack of band gap, the dimension of graphene sheet has to be reduced. After all, new physics (quantization) emerge when dimensions of materials are reduced. An infinite 2D graphene  could become a one dimensional plus quantization along one other direction, the consequence is the opening of a gap in the its dispersion spectrum. The resulting material from the 2D infinite sheet form a graphene nanoribbon (GNR). Depending on the nature of the ribbon edges, one gets two symmetry groups; Armchair Graphene Nanoribbon (aGNR) and Zigzag Graphene Nanoribbon (zGNR). Electron dynamics of both aGNR and zGNR have different properties, mostly due to the berry phase and pseudo spin \cite{KSasaki}. Edge states have significant contribution to graphene properties, because in a nanometer size ribbon, massless Dirac fermions can reach the edges within a femto-second before encountering any other lattice effects, like electron-electron interaction, electron-phonon interaction, etc.

In this paper, we study the phenomenon of negative differential conductivity (NDC) in GNRs. In conventional semiconductor devices, a negative differential conductive behavior is known to offer great potential for high frequency applications as Bloch oscillators, frequency multipliers, and fast switching devices. For this reason, the NDC effect has been greatly explored and discussed in several graphene nanostructures, particularly in \cite{VRyzhii}. NDC can also be observed in other graphene allotropes; carbon nanotubes (CNT) \cite{SSAbukari2}. The unique energy spectrum of holes and electrons in GNRs, especially its narrow gapless nature leads to nontrivial features such as Negative Differential Conductivity (NDC) in the THz regime \cite{VRyzhii}. 

In fact, we must emphasize that the phenomenon of NDC and Bloch oscillations in a material is a possible signature for THz generations in the material \cite{VRyzhii2,PKim}, since NDC occurs in the THz spectral range. Most of the methods of producing THz frequencies are experimental and only very few analytical (without computer numerics) approaches are known. Though Green function techniques have also been employed in some cases. Motivated by the fact that a rigorous analytical approach is necessary for studying NDC in GNRs, we adopt a semi-classical method used in references \cite{SSAbukari,SSAbukari2} for armchair and zigzag CNTs. We predict that GNRs should also reproduce similar results as in reference \cite{SSAbukari} because both CNTs and GNRs have almost the same full tight binding (TB) nonlinear complex band structure. 

The rest of this paper is organized as follows; In Ssection \ref{sec:The-Theory}, we derived the current density of aGNR and zGNR and imposed certain conditions to reduce the equations to forms appropriate for our models. The results obtained in Section \ref{sec:Results} are plotted and discussed in Section \ref{sec:Discussions}. The paper finally concludes in Section \ref{sec:Conclusion} where some recommendations for future applications are made.

\section{\label{sec:The-Theory} The theory}
As it is usually done in semi-classical treatment of quantum systems, we assume that the dynamics of the free $\pi$-electrons in graphene satisfies the time-dependent Boltzmann transport equation (BTE) in zero magnetic field. That is, 
\begin{equation}
	\frac{\partial f(k,t)}{\partial t}+\frac{eE(t)}{\hbar}\frac{\partial f(k,t)}{\partial k}=\Gamma[f(k,t)-f_{0}(k)].
	\label{eq:one}
\end{equation}
We are also assuming relaxation time approximation (RTA) and a spatial uniform graphene nanoribbon. Also, the inverse of the relaxation time $\Gamma$ is momentum independent. For the case of energy varying $\Gamma$, see \cite{MRabiu}. In Eq.\ref{eq:one}, $f_{0}(k)$ and $f(t,k)$ are the equilibrium and non-equilibrium Fermi electron distribution functions, respectively. $e$ is the electronic charge, $k$ is the electron wave vector and $\hbar$ is the reduced plank constant. We consider an external applied field
\begin{equation}
	E(t) = \sum_{j=0}^{n}E_{j}e^{i(\omega_{j}t+\alpha_{j})}\label{eq:two}
\end{equation}
as a superposition of $n$ harmonic waves polarized along one direction with the angular frequency $\omega$. The phase difference between the ($j+1$)th and $j$th wave being $\alpha_{j+1}-\alpha_j=\alpha$ is arbitrary, $j$ is an integer. $E_j$ are the field amplitudes. We require that $\omega_0=\alpha_0=0$. In the following section we will look at current density for GNRs.

\subsection{\label{sec:aGNR} Armchair and zigzag nanoribbon band structures}
The energy band structure of aGNR and zGNR is characterized by three parameters; band index $\lambda$, phase $\theta$ and wave vector $k$ \cite{KSasaki, FHipolito}. For aGNR
\begin{equation}
	\mathcal{E}^{\lambda}(k,\theta) =\lambda\gamma_0\sqrt{1 + 4cos^2(s\Delta\theta) + 4cos(s\Delta\theta)cos(kl)},\label{eq:threea}
\end{equation}
and for zGNR 
\begin{equation}
	\mathcal{E}^{\lambda}(k,\theta) =\lambda\gamma_0\sqrt{1 + 4cos^2(kl') + 4cos(s\Delta\theta)cos(kl')}.\label{eq:threeb}
\end{equation}
$\lambda=\pm1$. (+) for conduction band and (-) for valence band. $l=\sqrt{3}a/2$ and $l'=a/2$, $a$ is the lattice spacing with value 0.246$nm$, $\gamma_0 \sim 3.0eV$ is the overlap integral and $\theta$ is the phase perpendicular to the quasi-momentum $\hbar k$. The 1BZ of aGNR is bounded by $kl = [-\pi/2,\pi/2]$ and the zGNR is $kl' = [0,\pi]$. $k$ is parallel to the edge and has translational symmetry along this direction. For aGNR, the transverse wave vector (phase) is quantized according to the rule \cite{KSasaki}, $\theta_s = s\Delta\theta$ with $\Delta\theta = \frac{\pi}{\mathcal{N}+1}$ and $s=1,2,\ldots, \mathcal{N}$. Unlike aGNR, the nature of transverse wave vector quantization is complicated in zGNR, depending on both $k$ and $\theta$ as $\Delta\theta_s = (\pi j + \Lambda(k,\theta))/(\mathcal{N} + 1)$. However, for simplicity we assume $\Lambda$ is constant, say $\pi/2$, so that $\Delta\theta_s=\frac{(2j + 1)\frac{\pi}{2}}{\mathcal{N} + 1}$. Except this little subtlety for zGNRs, all that will be discussed in the following for aGNR are equally applied to the zGNR.

Now, employed translational invariance of the graphene ribbon in the reciprocal space and expand in Fourier series functions $f$, $f_0$ and $\mathcal{E}$ along the edge having the periodicity in $k$. i.e,
\begin{equation}
	f_0(k,\theta) = \sum_{r\neq 0}f_r(\theta)e^{irkl},\label{eq:five1}
\end{equation}
\begin{equation}
	f(k,\theta, t) = \sum_{r\neq 0}f_r(\theta)e^{irkl}\Phi_r(t),\label{eq:five2}
\end{equation}
\begin{equation}
	\mathcal{E}(k,\theta) = \gamma_0\sum_{r\neq 0}\mathcal{E}_r(\theta)e^{irkl}.\label{eq:five3}
\end{equation}
The Fourier coefficient $f_r$ is expressed as $f_r(\theta) = \sum_{s=1}^{\mathcal{N}}f_{rs}\Delta\theta\delta(\theta_s-s\Delta\theta)$ with 
\begin{equation}
	f_{rs} = \frac{l}{\pi s\Delta\theta}\int_{-\pi/2l}^{\pi/2l}dkf_0(k,\theta)e^{-irkl},\quad f_{rs} = f^*_{-rs} \label{eq:six1}
\end{equation}
and
\begin{equation}
	\mathcal{E}_{r} = \frac{l}{2\pi\gamma_0}\int_{-\pi/2l}^{\pi/2l}dk\mathcal{E}(k)e^{-irkl},\qquad \mathcal{E}_{r} = \mathcal{E}^*_{-r}. \label{eq:six2}
\end{equation}
$s$ in Eq.\eqref{eq:six1} counts the number of dimers $\mathcal{N}$ in GNRs. The factor $\Phi_r$ in Eq.\eqref{eq:six2} is a central point in this paper and so has to be determined. $r$ is an integer and not equal to zero. We consider a classical limit in which energy levels could be excited due to thermal fluctuations, i.e $\Delta\mathcal{E} << K_BT << \mathcal{E}_C$. This condition is also necessary for large enough field, so that charge carriers can escape low energy scattering \cite{SKSekwao}. The energy level spacing $\Delta\mathcal{E} = \pi W\gamma_0l/A$, $\mathcal{E}_C$ is the charging energy, $K_B$ is the Boltzmann constant, $T$ is the lattice temperature and $W$ is the graphene width. In what follows next, we will find the form of $\Phi(t)$. To do this Eqs.\eqref{eq:five1}, \eqref{eq:five2}, \eqref{eq:five3} are substituted in Eq.\eqref{eq:one} to yield 
\begin{equation}
	\frac{d\Phi_r(t)}{dt} = [\Gamma + ir\Omega(t)]\Phi_r(t)-\Gamma = 0, \label{eq:seven}
\end{equation}
where $\Omega(t) = \Omega_0 + \frac{el}{\hbar}\sum_{j=1}^{n}E_je^{i\omega_jt + \alpha_j}$ is the modulation degree of anharmonicity in electron motion. The solution of eq.\eqref{eq:seven} is a straight forward one,  using the boundary conditions $t=0$, $\Phi_r(0)=1$, one has 
\begin{equation}
	\Phi_r(t) = \frac{\Gamma\int dt e^{\Gamma t + i\sum_{j=1}^{n}\beta_je^{i\omega_jt + \alpha_j} + i\beta_0t}}{e^{\Gamma t + i\sum_{j=1}^{n}\beta_je^{i\omega_jt + \alpha_j} + i\beta_0t}}, \label{eq:eight}
\end{equation}
$\beta_j = erlE_j/\hbar\omega_j$ and $\beta_0 = \Omega_0 = erlE_0$. One can introduce product notation in eq.\eqref{eq:eight} as 
\begin{eqnarray}
	\mathcal{R}e\Phi_r(t) &=& \Gamma\prod_{j'=1}^{n}\prod_{j=1}^{n}\left[e^{-\Gamma t -i \beta_j' cos(i\omega_j't + \alpha_j') - i\beta_0t}\right]\nonumber\\
	&&\times\int dt e^{\Gamma t + i\beta_jcos(i\omega_jt + \alpha_j) + i\beta_0t},\label{eq:nine1}
\end{eqnarray}
\begin{eqnarray}
	\mathcal{I}m\Phi_r(t) &=& \Gamma\prod_{j'=1}^{n}\prod_{j=1}^{n}\left[e^{-\Gamma t -i \beta_j' sin(i\omega_j't + \alpha_j') - i\beta_0t}\right]\nonumber\\
	&&\times \int dt e^{\Gamma t + i\beta_jsin(i\omega_jt + \alpha_j) + i\beta_0t}.\label{eq:nine2}
\end{eqnarray}
Eqs.\eqref{eq:nine1}, \eqref{eq:nine2} are connected with the well known Bessel functions via Jacobi-Anger expansion
\[
	e^{\pm i\beta_jsin\theta} = \sum_{m=-\infty}^{\infty}J_m(\beta_j)e^{\pm im\theta},
\]
\[
 	e^{\pm i\beta_jcos\theta} = \sum_{m=-\infty}^{\infty}i^mJ_m(\beta_j)e^{\pm im\theta}.
\]
$J_m(\beta)$ is the $m^{th}$ order Bessel function. Using the expansions above, Eq.\eqref{eq:nine2} becomes

\begin{eqnarray}
\Phi_r(t) &=& \Gamma\sum_{n_{j'}=-\infty}^{\infty}\sum_{m_j=-\infty}^{\infty}\prod_{j'=1}^{n}\prod_{j=1}^{n}J_{n_{j'}}(\beta_{j'})\nonumber\\ 
&&\times\,e^{-\Gamma t -i \beta_{j'} sin(i\omega_{j'}t + \alpha_{j'}) - i\beta_0t}\nonumber\\
&&\times\, \int dt J_{m_j}(\beta_j)e^{\Gamma t + i\beta_jsin(i\omega_jt + \alpha_j) + i\beta_0t} 
\end{eqnarray}
Applying the integration in the preceding equation and letting $j=j'=1,2,3,\ldots$ and $m_j,\,n_j=\pm1,\pm2,\pm3,\ldots$
\begin{equation}
\Phi_r(t)=\sum_{n_j,\,\nu_j=-\infty}^{\infty}\prod_{j=1}^{n}J_{n_j}(\beta_j)J_{n_j-\nu_j}(\beta_j)\frac{e^{i\nu_j\omega_jt + i\nu_j\alpha_j}}{1+i\tau(\beta_0+\nu_j\omega_j)},
\end{equation}
where $\nu_j=n_j-m_j$.

\subsection{Sheet current density}

The sheet current density of graphene can be determined from the relation 
\begin{equation}
	j(t) = \frac{g_sg_v}{A}\sum_kev_kf_k. 
\end{equation}
The sheet area $A=WL$, with $W$ the width. $g_s$, $g_v$ are the spin and valley degeneracies respectively. For the aGNR,
\begin{equation}
	j(t) = \frac{g_sg_ve}{4\pi^2}\sum_{s=1}^{\mathcal{N}}\int\, dkv(k,\theta_s)f(k,\theta_s, \Phi_r).
\end{equation}
The velocity of Dirac fermions in graphene is defined as $v(k)= \partial\mathcal{E}/\hbar\partial k$. In terms of the Fourier coefficients,
\begin{equation}
	v(k,\theta) = \frac{i\gamma_0l}{\hbar}\sum_{r\neq 0}r\mathcal{E}_{rs}e^{irkl}
\end{equation}
giving
\begin{equation}
	j(t) = i\sum_{r=1}^{\infty}j_{0,r}\Phi_{r}(t) + c.c\label{eq:ja0}
\end{equation}
with
\[
	j_{0,r} = \frac{2g_sg_ve\gamma_0}{\pi l\hbar}\Delta\theta\sum_{s=1}^{n}r\mathcal{E}_{rs}f_{rs},\quad \quad j_{0,r}^* = -j_{0,-r}.
\]
Substitute $\Phi_r$ in Eq.\eqref{eq:ja0} to get
\begin{eqnarray}
	j(t) &=& i\sum_{r = 1}^{\infty}j_{0,r} \sum_{n_j,\,\nu_j=-\infty}^{\infty}\prod_{j=1}^nJ_{n_j}(\beta_j)J_{n_j-\nu_j}(\beta_j)\nonumber\\
	&&\times\,\frac{e^{i\nu_j\omega_jt + i\nu_j\alpha_j}}{1+i\tau(\beta_0 + n_j\omega_j)} + c.c.\label{eq:ja}
\end{eqnarray}
Note the $r$ dependence of $\beta_j$, $\beta_0$ and the summation over the index. Using the formalism by Litvinov and Manasson \cite{LitvinovM}, Eq.\eqref{eq:ja} can be put in a taylor-like expansion of $E_j$. i.e
\begin{equation}
	j(t) = j_{dc} +  \frac{1}{2}\sum_jE_j\sum_{\nu_j \neq 0} \sigma_{n_j\omega_j}e^{i\nu_j\omega_jt} + c.c + \cdots,
\end{equation}
where
\begin{equation}
	j_{dc} = \sum_{r=1}^{\infty}j_{0,r}\sum_{n_j=-\infty}^{\infty}\prod_{j=1}^nJ_{n_j}^2(\beta_j)\frac{i + \beta_0\tau + n_j\omega_j\tau}{1+[\tau(\beta_0 + n_j\omega_j)]^2} + c.c\label{eq:dc0}
\end{equation}
is the differential dc conductivity (for $\nu_j = 0$), and
\begin{eqnarray}
	\sigma_{n_j\omega_j} &=& 2\sum_{r=1}^{\infty}j_{0,r}\sum_{n_j=-\infty}^{\infty}\prod_{j=1}^n\frac{J_{n_j}(\beta_j)J_{n_j-\nu_j}(\beta_j)}{E_j}\nonumber\\
	&&\times \frac{i + \beta_0\tau + n_j\omega_j\tau}{1+[\tau(\beta_0 + n_j\omega_j)]^2}e^{i\nu_j\alpha_j} + c.c\label{eq:dcw}
\end{eqnarray}
is the large-signal dynamic nonlinear conductivity at $\nu_j$ harmonic with drive frequency $\omega_j$.

\section{\label{sec:Results} Negative differential conductivity}

\subsection{Pure dc limit}
To see immediately that Eq.\eqref{eq:dc0} demonstrates NDC, we consider a pure dc limit where $\omega_j \to 0$. The Bessel functions except the $n = 0$ term will vanish. The real part of the differential conductivity $\sigma (0) = \lim_{\omega_j\to 0}\partial j/\partial E_0$  becomes
\begin{equation}
	\sigma(0) = \sum_{r=1}^{\infty}\sigma_{0r}\frac{el\tau}{\hbar}\frac{1-(\beta_0\tau)^2}{1+(\tau\beta_0)^2}, \label{eq:NDCa}
\end{equation}
so that if $\beta_0 > \tau^{-1}$, the differential conductivity is negative and NDC is manifest in GNRs.

Electron dynamics may be come more complicated in the presence of high-frequency components in addition to the static electric fields. High negative differential conductivity thus may result in GNRs if an external drive force is applied. 

\subsection{Monoharmonics}

\begin{figure*}[thb!]
	\centering{\includegraphics[height=3.25in,width=6.5in]{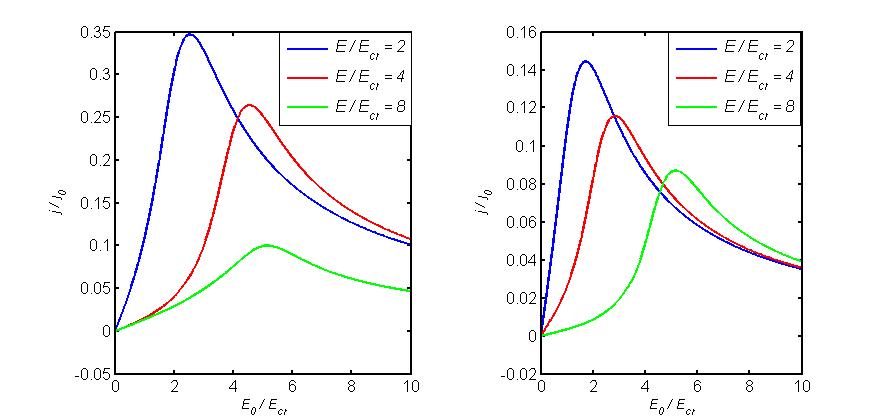}}
	\caption{Negative differential conductivity for (left) armchair and (right) zigzag graphene nanoribbons at different ac field amplitudes. $\omega\tau = 0.1$. The onset of NDC is at low static fields around $E_0 \sim E$. It departs from this condition at high fields}
	\label{fig:ANDC}
\end{figure*}

If one component of an ac filed in Eq.\eqref{eq:one} is applied, then $n = 1$ and Eq.\eqref{eq:dc0} simplifies to
\begin{equation}
	j = \sum_{r}j_{0r}\sum_{n=-\infty}^{\infty}J_{n}^2(\beta)\frac{\beta_0\tau+n\omega_1\tau}{1+(\beta_0\tau + n\omega\tau)^2}, \label{eq:NDCb}
\end{equation}
after dropping the subscripts on $n$. Here, it not clear immediately how NDC can be seen. To observe it, we plot $j$ versus $E_0$ for $\omega\tau << 1$ is shown in Fig.\ref{fig:ANDC} (left) for armchair ribbon and in Fig.\ref{fig:ANDC} (right) for zigzag ribbon,


\subsection{Biharmonics}

One can also allow the graphene nanoribbon subject to two ac fields. In that case, we let $j = 1, 2$ in our general formalism in Section \ref{sec:The-Theory} . This case has been studied in literature, especially in \cite{KNAlekseev} for superlattices. Eq.\eqref{eq:ja} takes the form
\begin{eqnarray}
	j &=& i\sum_{r = 1}^{\infty}j_{0r} \sum_{n_1,\,n_2=-\infty}^{\infty}\sum_{\nu_1,\,\nu_2=-\infty}^{\infty}J_{n_1}(\beta_1)J_{n_1-\nu_1}(\beta_1) \nonumber \\
	     && \times J_{n_2}(\beta_2)J_{n_2-\nu_2}(\beta_2)\frac{e^{i\nu_1\alpha_1 + i\nu_2\alpha_2}}{1+i\tau(\beta_0 + n_1\omega_1 + n_2\omega_2)}.\label{eq:ja1}
\end{eqnarray}
from which Eq.\eqref{eq:dcw} follows. We have eliminated the time dependence by averaging over the period of the fields to find the time-independent current $j$. In the left hand side, we replaced $\langle j(t) \rangle = j$, and in the right hand side a delta function emerges which ensures that $\nu_1 = -\frac{\omega_2}{\omega_1}\nu_2$. If $\omega_1 = \omega_2$, then one must put $\alpha_1 = 0$ and $\alpha_2 = \alpha$ so that $\nu_1 = \nu_2$. However, we shall generalized this to a case of commensurate frequencies. We exemplified the case by biharmonic having frequencies which can be periodic $\omega_2 = \mu\omega_1$ or non-periodic, $\omega_2 \neq \mu\omega_1$ with $\mu = 1,2,\ldots$. These two cases were studied in \cite{KSeeger, KAPronin} for semiconductor superlattices. Defining $j_{0,r} = \frac{2g_vg_se\gamma_0}{\sqrt{3}\hbar(n+1)a}\sum_{s=1}^nr\mathcal{E}_{rs}f_{rs}$, Eq.\eqref{eq:ja1} assumes the form
\begin{eqnarray}
j &=& i\sum_{r=1}^{\infty}j_{0,r}\sum_{n_1,\,\nu_2=-\infty}^{\infty}e^{i\nu_2\alpha}\nonumber\\
	 && \times\, \frac{J_{n_1}(\beta_1)J_{n_1+\mu\nu_2}(\beta_1)J_{n_2}(\beta_2)J_{n_2-\nu_2}(\beta_2)}{1+i\tau(\beta_0 + [n_1 + \mu n_2]\omega_1)}.
\end{eqnarray}
Simplifying further, we linearize with respect to one of the field amplitudes (say, $E_2$). For a week field, $\beta_2 << 1$ and $J_{n}(\beta) \approx (\beta/2)^2/n!$, which allows us to take $n_2$ ($n_2 - \nu_2) = \pm 1$ ($0,\pm1$). We obtained
\begin{eqnarray}
         j &=& i\sum_{r=1}^{\infty}j_{0,r}\sum_{n_1=-\infty}^{\infty}e^{in_2\alpha}\nonumber\\
	 & \times&\,\frac{J_0(\beta_2)\sum_{n_2=\pm1}J_{n_2}(\beta_2)J_{n_1}(\beta_1)J_{n_1 + \mu n_2}(\beta_1)}{1+i\tau(\beta_0 + [n_1 + \mu n_2]\omega_1)}, \label{eq:ja2}
\end{eqnarray}
with $\beta_{1,2} = erlE_{1,2}/\hbar\omega_{1,2}$. Where $l = \sqrt{3}a/2$ for armchair and $l' = a/2$ for zigzag graphene nanoribbons respectively. Finally, the current density becomes
\begin{eqnarray}
	j &=& \frac{el\tau^2}{\mu\hbar}E_2cos\alpha\sum_{r=1}^{\infty} j_{0r}\frac{\beta_0\tau + n\omega\tau}{1 + (\beta_0\tau + n\omega\tau)^2}\nonumber\\
		&&\times J_{n}(r\beta_1)\left\{J_{n - \mu}(r\beta_1) - J_{n + \mu}(r\beta_1)\right\},\label{eq:ja3}
\end{eqnarray}
\begin{figure*}[thb!]
\centering{\includegraphics[height=3.35in,width=6.75in]{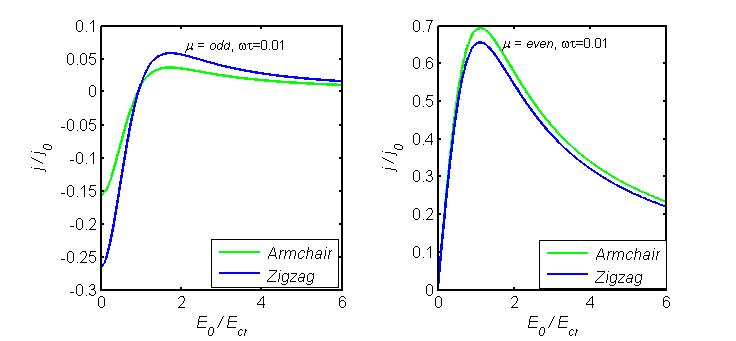}}
	\caption{Negative differential conductivity due to wave mixing of two ac field amplitudes. (left) $\mu = 2$ and (right) $\mu = 3$. The parameters used are $E_1 = 0.2E_{cr}$, $E_2 = E_{cr}$ and $\omega\tau = 0.01$.}
	\label{fig:AZNDC}
\end{figure*}
which reduces to the monoharmonic case when $\mu=1$.

\begin{figure}[thb!]
	\centering
	  \includegraphics[scale=.48]{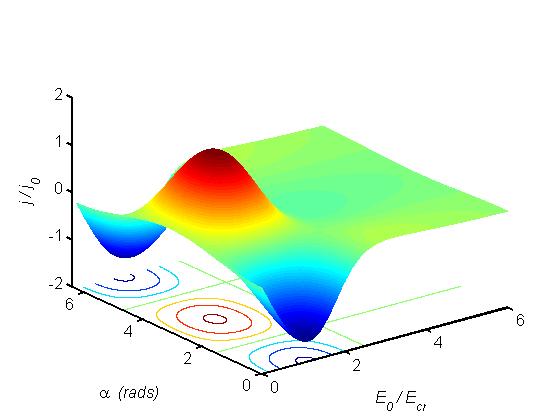}
	\caption{NDC of AGNR for simultaneously varying ac field amplitude and phase shift.}
	\label{fig:new3D}
\end{figure}

The nature of the NDC is observed for a simultaneously varying harmonic field and phase difference in a three dimensional plot shown in Fig.\ref{fig:new3D}

\section{ \label{sec:Discussions} Discussion}

In Fig.\ref{fig:ANDC}, normalized current density $j/j_0$ is plotted against reduced static electric field $E_0/E_{cr}$ for aGNR (left) and zGNR (right) for an applied ac field. At low fields up to $E_0 ( j_{max} )$, the quantum derivative of the $j-E_0$ characteristic yields a positive slope. A negative slope results for $E_0 > E_0 ( j_{max} )$. The whole of the region $E_0 > E_0 ( j_{max} )$ gives what is called Negative Differential Conductivity (NDC). A consequence of NDC in GNRs is a formation of electric field domains that impedes a continuous motion of electric field waves and thus blocks high frequency generation in these nanoribbons. NDC disappears quite faster in aGNR as $E \to \infty$ as compared to zGNR which is more rubust at this limit.

The curves in Fig.\ref{fig:AZNDC} demonstrate NDC, they are obtained at wave mixing of two commensurate frequencies, $\omega_2 = \mu\omega_1$. Fig.\ref{fig:AZNDC} (left) $\mu = odd$ and Fig.\ref{fig:AZNDC} (right) $\mu = even$. The onset of NDC in {\it odd}-harmonics occurs around $E_0 \sim E_1$, and in {\it even}-harmonics it starts at $E_0 \leq E_1$. In both cases, as in the previous NDC graphs, $\omega\tau << 1$.

The combined effect of phase shift and ac amplitude on NDC is depicted in Fig.\ref{fig:new3D}. There are three peaks at low bias fields at points ($E \sim E_{cr}, \alpha=0$), ($E \sim E_{cr}, \alpha=\pi$) and ($E \sim E_{cr}, \alpha=2\pi$). For now, it is not clear what these crests and throughs represents, they might be associated with field domains along one direction (for $\alpha = 0,2\pi$) and others along the opposite direction (for $\alpha = \pi$) and vice-versa.

\section{\label{sec:Conclusion} Conclusion}

We have demonstrated that graphene nanoribbons exhibit NDC regions in its $j-E_0$ characteristics at low bias field when $\omega\tau << 1$. NDC is observed either in the presence of bias field alone or by superimposing ac field amplitudes on the bias field. For one ac field, NDC occurs around $\omega\tau \sim 0.1$. When two ac fields at commensurate frequencies are applied, high-harmonic NDC emerge for both even and odd harmonics at rather very low frequencies $\omega\tau \sim 0.01$. The even-series gives pronounced high-harmonic NDC than the odd-series. The presence of high-harmonic NDC means that it is possible for high-frequency generation in graphene nanoribbons when electric field domains are suppressed at high enough applied frequencies $\omega\tau >> 1$ and $E_0 > E_{cr}$. We therefore suggest this approach for the study of terahertz generation in graphene.


\end{document}